\documentclass[12pt,preprint]{aastex}
\shorttitle{Temporal distribution of Cosmological Transients} \shortauthors{Howell et al.}

\begin{document}

\title{An improved method for estimating source densities using the temporal distribution of Cosmological Transients}

\author{E. Howell\altaffilmark{1}, D. Coward\altaffilmark{1}, R. Burman\altaffilmark{1} and D. Blair\altaffilmark{1}}
\affil{School of Physics, University of Western Australia, Crawley WA 6009, Australia}
\email{ejhowell@physics.uwa.edu.au} \altaffiltext{1}{School of Physics, University of Western Australia, Crawley WA
6009, Australia}

\newpage

\begin{abstract}
It has been shown that the observed temporal distribution of transient events in the cosmos can be used to constrain
their rate density. Here we show that the peak flux--observation time relation takes the form of a power law that is
invariant to the luminosity distribution of the sources, and that the method can be greatly improved by invoking time
reversal invariance and the temporal cosmological principle. We demonstrate how the method can be used to constrain
distributions of transient events, by applying it to \emph{Swift} gamma-ray burst data and show that the peak
flux--observation time relation is in good agreement with recent estimates of source parameters. We additionally show
that the intrinsic time dependence allows the method to be used as a predictive tool. Within the next year of
\emph{Swift} observation, we find a 50\% chance of obtaining a peak flux greater than that of GRB 060017 -- the highest
\emph{Swift} peak flux to date -- and the same probability of detecting a burst with peak flux $>$ 100 photons
$\mathrm{s}^{-1} \mathrm{cm}^{-2}$ within 6 years.
\end{abstract}





\newpage

\section{INTRODUCTION}

The brightness distribution of cosmological sources is conventionally used to constrain the luminosity function of the
sources, their evolution in density \citep{Peebles} and, for transient sources, their rate density
\citep{schmidt01,sethi01,totani97}. This method is applicable both to long-lived sources such as galaxies and to
transient events such as supernovae and gamma-ray bursts (GRBs). Estimates are obtained by fitting the number --
brightness distribution to models that include luminosity, source density and evolution effects. In the case of
transient events an additional parameter is available -- the event arrival times.

The temporal distribution of transient astrophysical populations of events has been described by the `probability event
horizon' (PEH) concept of \cite{coward05a}. This method establishes a temporal dependence by noting the occurrences of
successively brighter events in a time series. By utilizing the fact that the rarest events will preferentially occur
after the longest observational periods, it produces a data set with a unique statistical signature.

Here we show that a well-defined observation-time dependence is an intrinsic feature of the source distribution of
events. Using Swift GRB data we  demonstrate how this property can be used to constrain source distributions. We start
by presenting an analytical derivation of the peak flux--observation time relation, $P(T)$, for sources which are
uniformly distributed in Euclidean space and then describe its extension to cosmological models
(\S\ref{section_derivation}). We derive a simple power-law relation for $P(T)$ that is invariant to the luminosity
distribution of events.

We then utilize the PEH technique to show how $P(T)$ data can be extracted from a distribution of peak fluxes (\S
\ref{section_enhancedPEH}). We show that the PEH method can be greatly improved by invoking time reversal invariance
and the temporal cosmological principle: for time scales that are short compared to the age of the Universe, a
distribution of independent events is invariant with respect to temporal direction and there is nothing special about
the time when we switch on our detector.

As a test, we apply the $P(T)$ relation to \emph{Swift} long-GRB data. We demonstrate that the technique can be used as
a probe of the event rate density and luminosity distribution of the sources (\S\ref{section_FittingTool}), and as a
tool to predict the likelihood of future high peak flux events (\S\ref{section_PredictiveTool}).

\vspace{-4.0mm}
\section{THE PEAK FLUX -- OBSERVATION TIME RELATION}
\label{section_derivation}
%\subsection{Log P -- Log T relation for a Euclidean Universe}

In this paper we will define an \emph{event} to be an astrophysical transient occurrence with a duration much less than
the period of observation. Examples are GRBs and gravitational wave burst sources such as coalescing compact binaries
or core-collapse supernovae.

Consider a distribution of events defined within a Euclidean space by an event rate density $r_{0}$ and a luminosity
function\footnote{We use here the luminosity function for GRB sources, $\phi(L)$, which includes a normalization
constant to ensure that it integrates to unity over the range of source luminosities. This means that $\phi(L)$ has
units of inverse luminosity--see for example, \cite{pm01}.} $\phi(L)$ $(L_{\mathrm{min}}\le L \le L_{\mathrm{max}})$.
The observed peak flux, or `brightness', distribution of events over an observation time $T$ is a convolution of the
radial distribution of the sources and their luminosity function. For peak fluxes (photons
$\mathrm{cm}^{-2}\mathrm{s}^{-1}$) between $P$ and $P + \mathrm{d}P$:

\begin{equation}
\mathrm{d}N(P) = 4\pi T \int_{L_{\mathrm{min}} }^{L_{\mathrm{max}} } \phi(L)dL
 \hspace{1mm}r_{0} r^{2}dr , \label{eq_diff_source_count_euc}\\
\end{equation}

\noindent with $r=(L/4\pi P)^{1/2}$. The total number of events observed in time $T$ with a peak flux greater than $P$
is given by :

%\begin{eqnarray}
%\begin{align}
\begin{equation}
% \hspace{5mm}  N(> F) = N(P')\mathrm{d}P'\nonumber\\
N(> P) = T \Delta\Omega \int_{L_{\mathrm{min}} }^{L_{\mathrm{max}} } \phi(L)dL  \int_{0}^{\sqrt{ L/4{\pi}
P}}\!\!\!\!r_{0}\hspace{0.5mm}r^{2}dr\,, \label{eq_NF1}
%\end{align}
%\end{eqnarray}
\end{equation}

\noindent where the average solid angle covered on the sky has been accounted for by $\Delta\Omega /4 \pi$. The upper
limit in the integration over $r$ is the maximum distance for which an event with luminosity $L$ produces a peak flux
$P$.

For $r_{0}$ and $\phi(L)$ independent of position, integrating over the radial distance yields:

\begin{equation}
N(> P) = \frac{T r_{0}\hspace{0.5mm} \Delta\Omega/4\pi }{3 \hspace{0.5mm}\sqrt{4 \pi}}\hspace{0.5mm}
P^{-3/2}\hspace{0.5mm}\int_{L_{\mathrm{min}} }^{L_{\mathrm{max}} } \phi(L)L^{3/2} dL\,.
 \label{eq_NF2}
\end{equation}

\noindent This is the familiar log $N$--log $P$ relation, \mbox{$N(>P) \propto P^{-3/2}$}, a power law independent of
the form of the luminosity function \citep{horack94}.

To introduce the temporal distribution of events, we note that, as the events are independent of each other, the
individual events will follow a Poisson distribution in time. Therefore, the temporal separation between events will
follow an exponential distribution, defined by a mean event rate $R(r) = r_{0}(4/3)\pi r^{3}$ for events out to $r$.
The probability for at least one event $>P$ to occur in a volume bounded by $r$ during an observation time $T$ at
constant probability $\epsilon$ is given by:

\begin{equation}\label{eq_peh}
\mathcal{P}(n \ge 1;R(r),T)=  1 - e^{R(r)T} = \epsilon \,.
\end{equation}

\noindent For this equation to remain satisfied with increasing observation time:

\begin{equation}\label{eq_eps}
N(>P) = R(r)T =  |\mathrm{ln}(1 - \epsilon)|. \\
\end{equation}

Equations (\ref{eq_NF2}) and (\ref{eq_eps}) for $N(>P)$ combine to give the relation for the evolution of brightness as
a function of observation time:

\begin{equation}
P(T)\! =\!\! \left(\frac{r_{0}\hspace{0.5mm}\Delta\Omega /4\pi\hspace{0.5mm} }{3 \hspace{0.5mm}\sqrt{4
\pi}\hspace{0.5mm}|\mathrm{ln}(1 - \epsilon)|}\right)^{\!\!2/3}\!\! \left[\int_{L_{\mathrm{min}}
}^{L_{\mathrm{max}}}\!\!\!\!\!\!\!\phi(L)L^{3/2}dL \right]^{2/3}\!\!\!\!\!\!T^{2/3}\!\hspace{0.5mm}.
\label{eq_logPlogT_euc}\\
\end{equation}

This relation shows that for a simple Euclidean geometry, a log $P$--log $T$ distribution will have a slope of 2/3,
independent of the form of the luminosity function. One can consider that changes in $r_{0}$ create a horizontal offset
in the log $P$--log $T$ distribution, while changes in the integrated luminosity create a vertical offset. However, the
slope is fixed by the 3-D Euclidean geometry.

We can use the log $P$--log $T$ relation to produce curves defining the probability, $\epsilon$, of obtaining some
value of peak flux, $P$, within an observation-time, $T$, for a given $r_{0}$ \mbox{and $\phi(L)$}.

For a cosmological distribution of sources, equation (\ref{eq_logPlogT_euc}) must be modified to allow for cosmic
evolution. A standard Friedman cosmology can be used to define a differential event rate, $\mathrm{d}R(z)$, in the
redshift shell $z$ to $z + \mathrm{d}z$. The luminosity and flux will be related through $z$ by a luminosity distance
$d_{\mathrm{L}}^{2}(z)$ (see for example \cite{coward05a} or \cite{pm01}). In this case, solving equation
(\ref{eq_eps}) numerically, with $P = L/4\pi d_{\mathrm{L}}^{2}(z)$, will yield the cosmological log $P$--log $T$
relation.

\vspace{-4.0mm}
\section{AN ENHANCED PEH FILTER}
\label{section_enhancedPEH}

To utilize the time domain, we use the probability event horizon (PEH) filter of \cite{coward05a} to produce $P(T)$
time data. The PEH filter is a tool that exploits the temporal information encoded in a time series of transient events
and works by recording successively brighter events in a time series. \cite{howell07} demonstrated that the unique
statistical signature of events filtered in this way could be exploited to obtain rate estimates of transient events.
However, the significant probability of a bright event occurring early in an observational period meant that only a
small fraction of data was used by the method. As a result, large uncertainties were obtained in the estimates. There
are however, two ways in which the amount of usable data can be increased.

Firstly, the temporal cosmological principle implies that the PEH signature of a transient population of events is
independent of when a detector is switched on. Secondly, time reversal invariance allows the PEH filter to be applied
to a data set in both temporal directions. Thus, a time series of events can be treated as a closed loop which can be
interrogated in both directions. The observational period is now defined as the total length of the loop. The start
time for the PEH analysis is now arbitrary so, without loss of generality, we can choose any start time. This allows
the PEH filter to be applied in such a way that the brightest event can be set as the final event in a series. This
ensures that the PEH filter is applied to the full data stream and the process can be repeated in each direction,
increasing the quantity of PEH data. We refer to these techniques as `from max' plus `time reversal' (FMTR). We show
below how FMTR increases the PEH sample and significantly improves the statistical resolution when applied to the
\emph{Swift} data.

\section{APPLICATION TO \emph{SWIFT} DATA} \label{section_FittingTool}
\begin{figure}
\includegraphics[width=94mm]{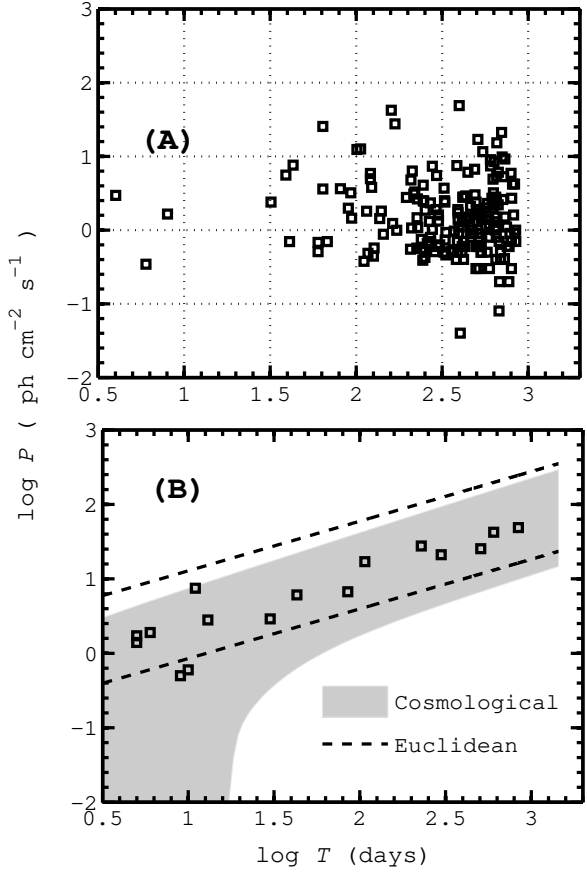}
\caption{Panel (A) shows the \emph{Swift} peak flux distribution as a time series. It is evident that as observation
time increases the probability of a bright event increases. Panel (B) uses a PEH filter to extract the geometrical
signature of the GRB distribution (shown as squares). Assuming an event rate of $0.1 \hspace{0.5mm}\mathrm{Gpc}^{-3}
\mathrm{yr}^{-1}$ \citep{guetta07}, the shaded area shows the cosmology dependent log $P$--log $T$ model corresponding
to a (5 -- 95)\% confidence band. The equivalent model for a Euclidean geometry is shown by the dashed curves -- the
two outliers, which have no associated redshifts, probably result from distant cosmological events.} \label{plotone}
\end{figure}

In this section, we will apply the log $P$--log $T$ relation to a cosmological population of long GRBs. To account for
the event rate and luminosity function, we will use estimates based on recent studies. Using the FMTR method, we will
extract a time-dependent sample of GRB peak fluxes from the \emph{Swift} data, and
demonstrate how it can be constrained by a log $P$--log $T$ fit.\\
\indent For our long-GRB peak flux sample we use data recorded by the \emph{Swift}\footnote{This data can be obtained
from the \emph{Swift} website http://swift.gsfc.nasa.gov/docs/swift/archive/grb\_table.html} satellite between 2004
December and 2007 April. We consider only bursts with confirmed peak fluxes detected within the 15--150 keV band of the
Burst Alert Telescope (BAT) and with $T_{90} \ge 2\hspace{0.5mm}s$ (a $T_{90}$ duration is the interval in which a
signal contains 90\% of its total observed counts). The total sample consists of
190 peak fluxes.\\
\indent Figure \ref{plotone}A displays the \emph{Swift} peak flux distribution of long GRBs as a time series. It is
apparent that as observation time increases, there is an greater probability of a bright event. By extracting
successively brighter events as a function of observation time, the PEH filter samples events from the low probability
tail of the distribution. The strong brightness--time dependence
of these events creates a unique statistical signature which can be modeled by the log $P$ -- log $T$ relation.\\
\indent In Table 1 we show the PEH filtered data. It is apparent how the time intervals between successive events
increase with observation time. This is a result of a progressive sampling of the rarer events of the
distribution.\\
\indent To apply the log $P$--log $T$ relation to the filtered data we must first set up a model to account for the
source rate evolution and luminosity distribution. We use a model from the recent study of \cite{guetta07}. They employ
a `flat-$\Lambda$' cosmological model and \mbox{$H_{0}=65$ km s$^{-1}$ Mpc$^{-1}$} for the Hubble parameter at the
present epoch. For the isotropic luminosity function of GRB peak luminosities, $\phi(L)$, they use a broken power law
form based on the work of \cite{schmidt01}. Assuming that the rate of GRBs traces the global star formation history of
the Universe, they employ a number of different star formation rate models. For each, they determine best fitted values
for the luminosity function and event rate density. For this study, we use their model (i) parameters, which are based
on the SF2 star formation rate model of \cite{pm01}. These parameters include fitted values for the luminosity function
and a local event rate density $r_{0} = 0.1\hspace{0.5mm} \mathrm{Gpc}^{-3} \mathrm{yr}^{-1}$.  To account for the
average solid angle covered on the sky by \emph{Swift}, we use a value of $\Delta\Omega =
1.33$ \citep{Band03}.\\
\indent Figure \ref{plotone}B shows a log $P$-- log $T$ fit to the PEH filtered data (shown by squares), using the
fitting parameters of \cite{guetta07}: i.e. there are no free parameters in the comparison between theory and
observation. We define a 90\% confidence band -- shown by the shaded area -- corresponding to the $\epsilon = 95\%$
(top) and $\epsilon = 5\%$ (bottom) probabilities of detecting at least one event within an observation time $T$. We
see that the data is well constrained. The fit shows that by using only a small sample of the brightest events, it is
possible to extract the geometrical signature of the source population and to test estimates of the luminosity
function and rate density of events.\\
\indent The dashed lines of Figure \ref{plotone}B show the 90\% confidence band corresponding to the Euclidean model
using equation (\ref{eq_logPlogT_euc}). We see that two of the first few events lie outside the Euclidean curves but
are constrained by the cosmological model. These events, occurring at early observation times, most likely result from
sources at large cosmological distances. The very bright events at late observation times are more probable -- it is
apparent that the Euclidean and cosmological curves begin to converge in this regime. To account for non-uniformly
distributed sources, the method could be refined to take into account the spatial distribution of potential host
galaxies.

\indent To test the power-law dependence in equation (\ref{eq_logPlogT_euc}), we have performed least-squares fitting
to the PEH filtered data using the Euclidean log $P$-- log $T$ curve as a linear regression model. By setting the power
as a single free parameter, we obtained a value of $0.67 \pm 0.02$, confirming the 2/3 slope derived in section
\ref{section_derivation}.

\vspace{-4.0mm}
\section{A PREDICTIVE APPLICATION OF THE LOG P -- LOG T RELATION} \label{section_PredictiveTool}

%\begin{enumerate}\item

\begin{figure}
\includegraphics[width=84mm]{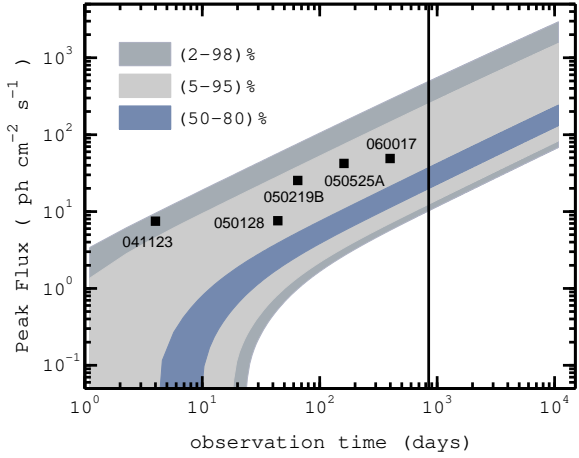}
\caption{The log $P$--log $T$ relation used as a predictive tool. The successive maximum peak fluxes detected by Swift
since the start of operation are shown by squares. The shaded areas show detection confidence bands corresponding to
different probability values (shown in the legend). The current \emph{Swift} observation time (398 days) is shown by
the solid line. The plot shows that after 3 days of operation, there was a $<5\%$ probability of detecting an event
with peak flux equal to the first in the PEH sample, GRB 041123. We see that within the next year, there is a 50\%
chance of obtaining a peak flux greater than for GRB 060017 -- the most intense burst (in peak flux) detected by Swift
-- and the same probability of obtaining a burst with peak flux $>$ 100 photons $\mathrm{s}^{-1} \mathrm{cm}^{-2}$
within 6 years. } \label{plottwo}
\end{figure}

Figure \ref{plottwo} illustrates how the log $P$--log $T$ relation can be used as a predictive tool. By mapping the
temporal evolution of detection probability, the maximum brightness of future events can be constrained
\citep{coward05b}. As in Fig. \ref{plotone}B, the shaded areas show the log $P$--log $T$ detection confidence bands
corresponding to different probability values (shown in the legend). The current \emph{Swift} observation time (398
days) is shown by the vertical solid line. For predictive applications it is essential that the true temporal sequence
of events is retained. Therefore, rather than using the FMTR technique, we apply the unmodified PEH filter from the
time of the first event. Comparing with Fig. \ref{plotone}B, we see that the FMTR method has increased the sample by
220\%, of which 80\% is gained by incorporating time reversal. Using 100 Monte-Carlo simulations, we find a mean
fractional increase in data of $\sim 35\%$, compared to the unmodified PEH filter. The larger than expected filter
output using \emph{Swift} data, implies that the FMTR method can be further optimally tuned.

The plot shows that after 3 days of operation, there was a $<5\%$ probability of detecting an event with peak flux
equal to the first in the PEH sample, GRB 041123. The Poisson probability of detecting at least one event within a Gpc
at this time is $\sim 2 \times 10^{-6}$. This implies that this event occurred at a considerable cosmological distance.
The event GRB 050525A is the brightest long GRB with a secure redshift, $z=0.606$ -- the Poisson probability of at
least one event within this volume is 35\%. The next brightest event, GRB 060017, is the most intense burst, in terms
of peak flux, detected by Swift.

As a demonstration of the predictive nature of the the log $P$ -- log $T$ relation, Figure \ref{plottwo} shows that
there is a 50\% probability of obtaining an event with a peak flux greater than that of GRB 060017 within the next year
and an 80\% probability within 5 years.

The curve predicts that there is a 50\% (80\%) probability of obtaining a burst with peak flux $>$ 60 photons
$\mathrm{s}^{-1} \mathrm{cm}^{-2}$ within 2 (8) years. To determine the feasibility of this prediction, we consider
again GRB 050525A, which had a peak flux of 42.3 photons $\mathrm{s}^{-1} \mathrm{cm}^{-2}$. Using this burst's
redshift and converting to a peak luminosity, we find that an equivalent burst would have to occur within $z \approx
0.5$ to produce a peak flux of this level. The Poisson probability of at least one event at this distance within the
next two years \mbox{is $\sim 90\% \hspace{0.5mm}(99\%)$}. If we consider a burst with a peak flux of 100 photons
$\mathrm{s}^{-1} \mathrm{cm}^{-2}$, we find that an event of this peak flux is 50\% probable within 6 years. Such an
event would correspond to a GRB 050525A-equivalent burst occurring within $z \approx 0.4$, for which the probability is
99\%.

The log $P$ -- log $T$ technique naturally uses the brightest events of a data set. As the probability of obtaining a
GRB afterglow increases with peak flux, the method can be used to predict the expected occurrence of events at low $z$.

\section{SUMMARY}
\label{section_conclusions}

We have provided a clear demonstration of the log $P$--log $T$ relation by applying the PEH filter to the \emph{Swift}
GRB peak flux distribution. A log $P$--log $T$ model with no free parameters was fitted to filtered data confirming the
power law $P \propto T^{2/3}$ in the Euclidean limit; the power law is independent of the form of the luminosity
distribution.

The FMTR method significantly improves the PEH method, which was previously disadvantaged by using only a small
fraction of a data stream. We have shown that FMTR enables the PEH filter to use over 8\% of the available data, making
it a practical tool for cosmology.

We have shown that the PEH technique can be used as a predictive tool. Comparing observation with prediction provides
an additional means to test rate estimates and evaluate source parameters such as the limits of the luminosity
distribution.

In a future study, we intend to apply the FMTR method to both the \emph{Swift} and \emph{BATSE} GRB data. We will
investigate the efficiency of the method in determining constraints on the rate density and limits of the luminosity
function.

\section*{ACKNOWLEDGMENTS}
E Howell and D Coward are supported by the Australian Research Council.
% This research is part of the research program of the Australian Consortium for Interferometric Gravitational Astronomy
%(ACIGA).
\vspace{-4.0mm}
\bibliographystyle{mn2e} % Choose Phys. Rev. style for bibliography
%\bibliography{swift_peh}        % qhe.bib is the name of our database
{}

\newpage

\begin{table}
\begin{centering}
\begin{tabular}[scale=1.0]{cccc}
  \hline
GRB            &Peak Flux    & Redshift & Observation Time\\
%GRB            &Peak Flux %(ph$\hspace{0.5mm}\mathrm{s}^{\!-1} \mathrm{cm}^{\!-2})$ %    & Redshift & Observation time (days)\\
               &(photons$\hspace{0.5mm}\mathrm{s}^{-1} \mathrm{cm}^{-2})$           &           & (days)        \\
  \hline
%060123       &0.04                     & $1.099 $     &  1 \\
060202       &0.5                     & $ $     &  8 \\
060203       &0.6                    & $$     &  9 \\
\hspace{2.0mm}060204B      &1.3                     & $ $     &  10 \\
060206       &2.8                     & $4.045 $     &  12 \\
\hspace{2.0mm}060223B      &2.9                    & $ $     &  29 \\
060306       &6.1                     & $ $     &  42 \\
060418       &6.7                     & $1.49 $     &  84 \\
\hspace{2.0mm}060510A      &17.0                    & $ $     &  106 \\
061121       &21.1                     & $1.314 $     &  297 \\
\hspace{2.0mm}050219B      &25.4                     & $ $     &  506 \\
\hspace{2.0mm}050525A      &42.3                     & $0.606 $     &  602 \\
060117       &48.9                     & $ $     &  839 \\
  \hline
%060111B      &  1.4                    &         &4\\
\hspace{2.0mm}060111A      &  1.72                   &         &4\\
060110       &  1.9                    &         &5\\
060105       & 7.5                     &         &10\\
060603       &27.6                     &2.821    &227\\
  \hline
\end{tabular}
\caption[waveform parameters]{Data extracted from the \emph{Swift} peak flux distribution of long GRBs using the PEH
filter. The observation time is determined by treating the time series as a closed loop and setting the last event in
the series to be the brightest burst. The lower set of data was obtained by invoking time reversal. A log $P$ -- log
$T$ fit to this data is shown in Fig. \ref{plotone}}
\end{centering}
\label{table1}
\end{table}

\newpage

\newpage

\end{document}